# Incrementally Introducing Process Model Rationale Support in an Organization

Alexis Ocampo[a], Jürgen Münch[a] and William E. Riddle[b]

[a]Fraunhofer Institute for Experimental Software Engineering,
Fraunhofer Platz 1, 67663 Kaiserslautern, Germany
{ocampo, muench}@iese.fraunhofer.de
[b]Solution Deployment Affiliates
223 N Guadalupe #313, Santa Fe, New Mexico, USA 87501
riddle@WmERiddle.com

**Abstract:** Popular process models such as the Rational Unified Process or the V-Modell XT are by nature large and complex. Each time that a new release is published software development organizations are confronted with the big challenge of understanding the rationale behind the new release and the extent to which it affects them. Usually, there is no information about what has changed or most importantly why. This is because of the lack of a flexible approach that supports organizations responsible for evolving such large process models in documenting their decisions and that reflects the extent of the capabilities to which they can provide this information. This paper describes an approach to incrementally deploying rationale support as needed to match an organization's needs, the capabilities and interests of the organization's process engineering teams, and the organization's willingness to support the effort required for the collection and application of the rationale information.

## 1. Introduction

Software process models support software engineers in systematically performing the engineering processes needed to develop and maintain software products. As these processes are enacted, suggestions and needs for adjustment or refinement arise, which, in turn, demands an evolution of the models. Changing these models in an organization is typically a complex and expensive task. In many cases, due to budget and time constraints, arbitrary decisions are made, and process models are evolved without storing or keeping track of the justification behind such changes. This frequently results in inconsistencies or ambiguity being introduced into the process models.

The work presented in this paper responds to the need for systematically performing changes to a process model by contributing an approach for rationale support of process model evolution called "REMIS". REMIS guides process engineers in the tasks of capturing the reasoning (i.e., rationale) behind such changes and analyzing the evolution. REMIS has been developed in a bottom-up fashion based on observations and experience from different case studies (industrial projects). The main contribution of REMIS to current research in the field of process model evolution consists of transferring and adapting design rationale concepts in order to support systematic process model evolution. Figure 1 shows the specific contributions which are: a) a conceptual model for describing the rationale for software process model changes, b) a method for capturing and analyzing the rationale for software process model changes, c) a classification of common situations for process model change, d) a tool that supports the method and e) an incremental deployment strategy. Previous publications describe in detail the contents of a), b), c) and d). This paper describes e), the method-deployment strategy.

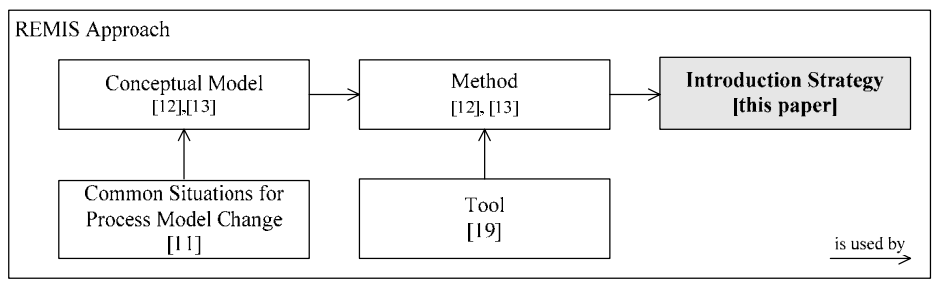

**Figure 1. The REMIS Approach**

Why is a strategy so important and necessary for introducing the REMIS approach into an organization? One special characteristic of rationale approaches is their degree of intrusiveness in the modeling process. That is, the extent to which the approach interferes with the modeling process. Such interference can happen not only during the capture of rationale but also during the retrieval of this rationale. The reason to highlight this characteristic is based on a long-term discussion about the costs and cultural implications of capturing rationale information in the rationale research and practitioner communities [5]. Although the approaches are in constant maturation, the resistance of practitioners to capture rationale information has been associated with intrusiveness. More intrusive approaches need a stronger accompanying process in order to be successful than less intrusive ones [17]. Therefore the deployment strategy presented in this paper is oriented toward mitigating this intrusiveness risk based on the assumption that a process engineering team must capture and apply information about process modeling decisions and their rationale according to the team's needs, capabilities and allocated-effort.

Two case studies used as input for the definition of the deployment strategy – the central part of this paper – are presented in the Section 2. This is followed, in Section 3, with a brief accounting of the requirements for a conceptual model, an associated rationale documentation method, and a deployment strategy addressing the problems revealed by the case study. Section 4 then discusses in brief the conceptual model and the method. Section 5 describes in detail the strategy for incrementally deploying the REMIS approach as needed by a process engineering team and as allowed by their capabilities. The paper ends first – in Section 6 – with a discussion of how the conceptual model, the method and deployment strategy satisfies the requirements given in Section 3, and then – in Section 7 – a summary of the work presented in this paper and a discussion of how the REMIS approach, its underlying conceptual model, and the deployment strategy might be improved through further research and development.

## 2. Case Studies for Eliciting Requirements on Rationale for Process Model Evolution

This section presents in brief, the experience captured in two different case studies in which rationale information was collected while evolving large and complex process models. A more detailed description of each case study, i.e., the study's definition, design and results can be found at [20]. The observations of the first case study served as inputs for its application in the second case study. The conceptual models used in both iterations and observations on the feasibility of their use, constituted the basis for the requirements of the REMIS approach and the deployment strategy.

**ESA Case Study**

For the European Space Agency (ESA), the relevant standards applicable for developing software are: ECSS-E-40B Space Engineering - Software [6] (mostly based on the ISO 12207 standard [8]), and ECSS-Q80-B: Space Product Assurance - Software [7]. Organizations or projects that are part of ESA are required to develop and use specific tailorings of the ECSS standards suited to their work. This is a particularly complex task because it requires detailed understanding of the whole standard, something that an average software developer or project manager usually does not have [16]. At the ESA Space Operations Center ESOC (the ESA organization where this project took place), this tailoring was called the Software Engineering and Management Guide (SEMG) [9] and was used for all their major projects.

After several years of experience with the ECSS standards, these were revised by ESA, and a new version was published. This also meant that the SEMG had to be revised, in order to be compliant with the revised ECSS standard. This compliance had to be proven by means of traceability of every ECSS requirement to its implementation and by providing a tailoring rationale for every tailored requirement.

*The goal of this study was to analyze the feasibility of the conceptual model and the approach for documenting and analyzing the rationale for process model change.*

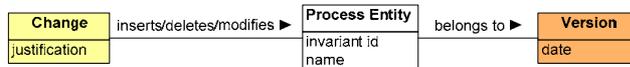

**Figure 2. Conceptual Model - ESA Case Study**

Figure 2 shows the conceptual model. *Changes* result from decisions captured in the justification and are performed on Process Entities. Some examples of Changes performed to Process Entities are: *Entity x has been inserted*; *Entity y has been deleted*; and *Entity x has been modified*. A *Process Entity* reflects a concept defined by a vocabulary/notation for modeling/describing process models, e.g., SPEM [14], V-Modell XT [10], SPEARMINT [4] and BPML [15].

The data about the changes to the SEMG were collected in meta-information tables attached to each section. Process engineers provided information about the rationale for a change each time a change was performed to the SEMG standard. Then they used an automated mechanism for storing this information in a database [12].

**Observations:** The tables that were used by process engineers for describing what changed and why were very useful for systematic reviews. However, sometimes the provided information about what changed was too detailed, sometimes too abstract. This might be due to the fact that the conceptual model did not anticipate a difference between finely granular changes (e.g., grammar errors or misspellings) and larger ones (e.g., wrong control flow). The lack of structure of a justification influenced the understandability of the collected information. The ESA reviewers commented on confusing justifications that identified <u>what</u> was performed instead of information on <u>why</u>. ESA reviewers also missed information concerning the alternatives taken into account by process engineers before performing the change. That information could have help reviewers understand faster the rationale and avoid unnecessary discussions. These findings motivated the need to change the conceptual model and the instrumentation and to use it in a second case study.

**ASG Case Study**

In this case study, process engineers were in charge of defining, establishing, evaluating, and systematically evolving the development process model applied in the project to develop a platform for Adaptive Services Grid (ASG) [1].

In general, the main idea behind the systematic evolution approach followed was to start with commonly accepted process knowledge, refine it with information gathered from the practitioners, and therewith improve the textual descriptions and diagrams of the process according to the real project needs. The conceptual model and the method developed in the ESA case study [2] were extended as well as the tool-assisted way of editing and storing the process model and its rationale information.

*The goal of this study was to analyze the feasibility of the refined conceptual model and the refined approach for documenting and analyzing the rationale for process model changes.*

As in the previous case study, meta-information tables were used as a means for realizing the conceptual model or the rationale for process model changes [12]. In practice, the process engineer discussed and resolved the issue while introducing the corresponding rationale information, then performed the changes to the respective process entities, and finally established a reference to the corresponding rationale concept.

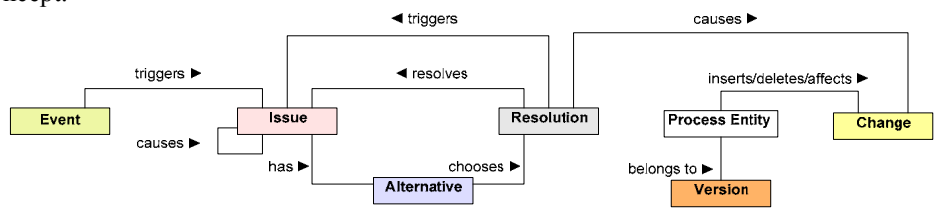

**Figure 3. Conceptual Model - ASG Case Study**

In the extended conceptual model (shown in Figure 3), an *Event* is a trigger of issues. Events may be characterized by a name and a short description (i.e., two of its attributes may be *name* and *short_description*). Events may also be characterized by their *type*. At least two types of events are possible: internal (e.g., new/updated corporate policies, e.g., policy changes stemming from changes to an organization's business goals) and external (e.g., new/updated software engineering technology, e.g., new testing support tools and techniques). *Issues* are situations that arise as a consequence of an Event, that need to be addressed, and that are related to a part of the described system. Additionally, an Issue may be categorized by its type. This type may be selected from a classification of issues pertinent to an organization. At this point, REMIS reflects a general, organization-independent classification of issues resulting from the ESA case study (i.e., imprecision; verbosity; inaccuracies; non-compliance; inconsistency).

Issues are often stated as *questions*. The question has the purpose of forcing software engineers to reason about the situation (the problem) they are facing before starting to think about or providing resolutions (the potential solutions). Some of the characteristics of an issue are a *synoptic_description*, a *status* (e.g., open, closed), and a *detailed_discussion*. The *detailed_discussion* may be used to capture the minutes of a meeting, E-mails, memos, letters, etc. in which the issue was discussed by software engineers or stakeholders.

*Alternatives* are assigned to an Issue; at least one Alternative might be proposed to resolve an Issue. Alternatives are described at least by a *subject*, and more comprehensively in a *description*. The assessment describes the acceptance of the alternative with respect to the characteristics pertinent to evaluating its achievement, e.g., its feasibility, cost and required-effort. Usual values are positive or negative. A *Resolution* changes the process model. A Resolution might lead to opening more Issues. Every Resolution is characterized by a *short_description*, a *long_description*, a *justification*, and a *status* (for example, open or closed). The justification is included to be able to capture a summary of the analysis of the different alternatives as well as a short note.

**Observations:** The extended conceptual model played an important role, because it allowed structuring better the reasoning behind a decision compared to the previous study. Especially concerning the alternatives taken into account. Having this information motivated self-reflection about the real need for changing the process model. Equally, the structure of the conceptual model allowed reusing this information in a straightforward way, before performing future changes. The types of events and issues provided a means for starting up a classification of common situations for process model change.

## 3. Requirements for a Rationale for Process Model Changes

The case studies revealed that the fundamental need is to collect rationale information which facilitates making and justifying design decisions underlying the process model's evolution in response to changes to its requirements or its operational context. This includes information about the alternatives which were considered and the rationale underlying the adoption or rejection of the various alternatives. Further, the case studies indicate that it is important that this information not merely captures low-level, "micro" details but that it be possible to integrate over the detailed information to provide information at the higher, "macro" pertinent to process model issues. Therefore, the basic requirement is:

- R1: Support the collection of information which may be directly used, or may be interpreted and analyzed, to understand alternatives, choose among them and justify their choice or rejection as necessary.

The case studies also indicate that the effort required to collect the information should be acceptable, as "minimal" as possible. Partially, this means that the planning of process model evolution activities account for the fact that some effort will be needed; an evolution plan must include an allocation of effort for collecting and applying rationale information. The information collection and application effort must be an acceptable increase over the effort needed for the evolution activities. Doubling the effort would obviously be unacceptable. Based on the authors' experiences, a 33% increase is probably an upper-bound, with the increase normally being in the range 15-20% with larger increases only when justifiable, for example when the system will undergo extensive independent review.

Accommodating a restriction such as this upon the information collection and application effort requires supporting the effort with at least guidelines – and, even better, guideline-implementing tools and techniques – that enhance the software engineers' abilities. It also requires the ability to customize the guidelines, tools and techniques to both enhance the pertinence, and therefore value, of the information and eliminate effort 'wasted on' the collection of unnecessary information.

This leads to two additional requirements:

- R2: Provide guidelines helping software engineers efficiently as well as effectively carry out rationale information collection and application tasks. When possible, provide tools (e.g., information templates) and techniques (e.g., decision making-support approaches) which implement the guidelines and reduce the effort needed to follow the guidelines.
- R3: Allow customization of a 'default' set of guidelines, tools and techniques serving to match the needs for specific process model evolution activities.

The final requirement is not directly revealed by the case studies. Rather, it comes from further considering the issue of customization. Requirement R3 reflects the need to customize with respect to the nature of the process model being evolved. The final requirement reflects a need to customize the guidelines, tools and techniques to match the abilities of an organization's software engineering, their tolerance for carrying out 'overhead' tasks, and the organization's willingness to support the extra effort needed to collect and apply rationale information (i.e., the organization's tolerance for effort increases needed to collect and apply the information). This requirement is:

- R4: Allow the incremental adoption of the guidelines, tools and techniques in steps of increasing scope, depth, difficulty, effort and value.

## 4. The REMIS Approach

The final conceptual model underlying REMIS, shown in Figure 4, results from the incremental, iteration-driven research strategy (based on the previously described case studies) aimed at understanding the information needs for capturing the rationale underlying change to a process model.

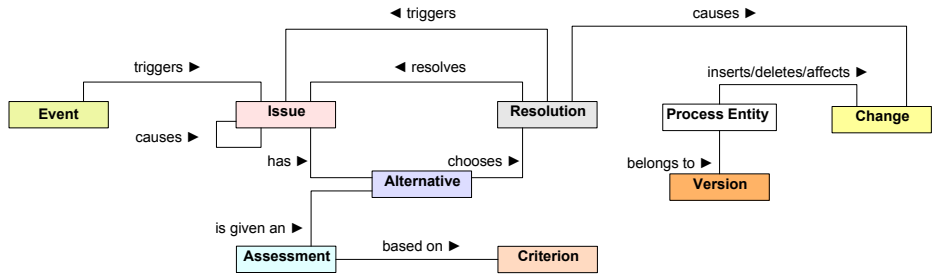

**Figure 4. REMIS Conceptual Model**

It can be seen in Figure 4 that compared to the conceptual model described in the ASG case study, that the following additional concepts have been introduced. An Alternative's *Assessment* is based on criteria. A *Criterion* is an influencing factor pertaining to a given organization in a certain context. A set of Criteria characterize the context in which changes are made. Criteria are important not only for comparatively assessing Alternatives, but also for recording evidence of the most influential factors that affect a decision. In the software design domain there is a noticeable lack of research regarding the Criteria affecting the assessment of design alternatives. For lack of a better approach, at this point in time the REMIS approach relies up the GQM paradigm [3] to dynamically, case-by-case, define the Criteria that affect an organization's software process modeling evolution efforts. This paradigm explicitly includes weights reflecting the importance of Criteria for an organization in a given evolution context with respect to other Criteria. Finally, every Resolution identifies changes which satisfy the various Criteria.

The method provided by REMIS (see Figure 5) is also based on experiences from the case studies and well supported by the conceptual model.

The following paragraphs discuss briefly the purpose and description of the method's product flow (for a more detailed discussion, please refer to [12]).

The purpose of the activity *Analyze change request* is to understand, assess, and prioritize the feedback provided by engineers or practitioners concerning the process model. Rationale visualization can be useful at this point for answering different types of questions relevant to process engineers. Examples of such questions are:

- Which still-open issues may conflict with the change/improvement proposal being analyzed?
- Which process model entities are affected by a previous resolution that conflicts with the new change/improvement proposal being analyzed?

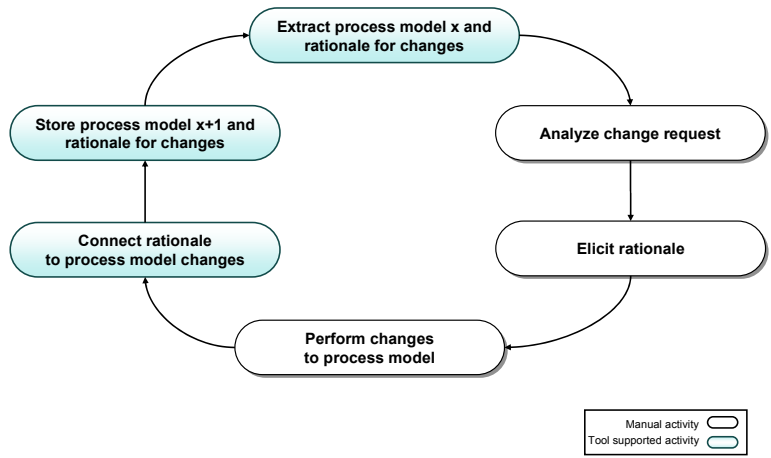

**Figure 5. REMIS Method**

The proposals are then prioritized. The process engineer selects those improvement proposals that should be considered according to the prioritization. Additionally, the process engineer decides whether the rationale should be elicited *synchronously (i.e., while performing the changes)* or rather *asynchronously (i.e., after performing the changes)*. The decision should normally be based on factors such as (a) relevance of the change/improvement proposal; (b) available resources; (c) available infrastructure; and (d) degree of maturity in eliciting the rationale.

The activity *Elicit rationale* consists of the process engineer analyzing and discussing with other stakeholders (e.g., project manager, quality manager) the change improvement proposals and deciding on a strategy for implementing the resulting changes. The reasoning behind the decision is captured during the analysis and/or discussions. Existing rationale information (that explains the evolution of the process model up to that moment) can support the process engineer in this activity. Rationale visualization can be useful again at this point for answering different types of questions.

Once the process engineer is sure about what changes to perform, she/he proceeds to the activities *Perform changes to model entities* and *Connect rationale to process model changes*. The process engineer then will implement the agreed-upon changes to the set of process model entities by using the specific process modeling tool used in his/her organization. In order to connect the rationale to the just performed process model changes, the process engineer can use two different techniques: one that mimics the technique used in the case studies and proposes inserting references to the rationale information directly into the process model entity being altered [12]. A second one that consists of after performing the changes (i.e., *asynchronously*) identifying the set of changed process model entities (by means of an special technique for comparing models called Delta-P [18]) and inserting a reference to the respective rationale for each one of those changes [13].

The purpose of the activity *Store new process model version x+1 and rationale for changes* is to make persistent the changes performed to a model and to annotate the model with a new version identifier. The process model evolution repository consists of a body of content formed by process model entity instances of a well-defined meta-model and the rationale information. The activities *Connect rationale to process model changes* and *Store new process model version x+1 and rationale for changes* are supported by the REMIS tool [19] in order to systematically keep the consistency between the different versions of the process models and its rationale.

## 5. The Incremental Introduction Strategy

The capture and visualization of the rationale for process model evolution must be accomplished in a systematic manner. Convincing an organization to change the way it works or to adapt to a new mechanism is a complicated task. Therefore – and based on the experiences of the case studies reported in [2], [12] and [13] – a staged incremental method, which facilitates the institutionalization of rationale and visualization into a software development organization, is proposed. This means that organizations have to incrementally learn how to collect rationale, what to collect, how to use it, and, especially, they have to identify which level of "maturity" in rationale-driven evolution they want to achieve. Figure 6 presents the different steps defined for incrementally introducing and institutionalizing rationale support for process model evolution. One advantage of using the RDF notation [21] as a basis for the specification of the conceptual model is the possibility of incrementally adding concepts to the rationale vocabulary. This facilitates gradual introduction as well as the design and implementation of tool support. The following paragraphs provide a more detailed description of the activity for eliciting process model rationale, highlighting the differences for different levels of deployment - i.e., REMIS 0, 1, 2 and 3 - identified in Figure 6.

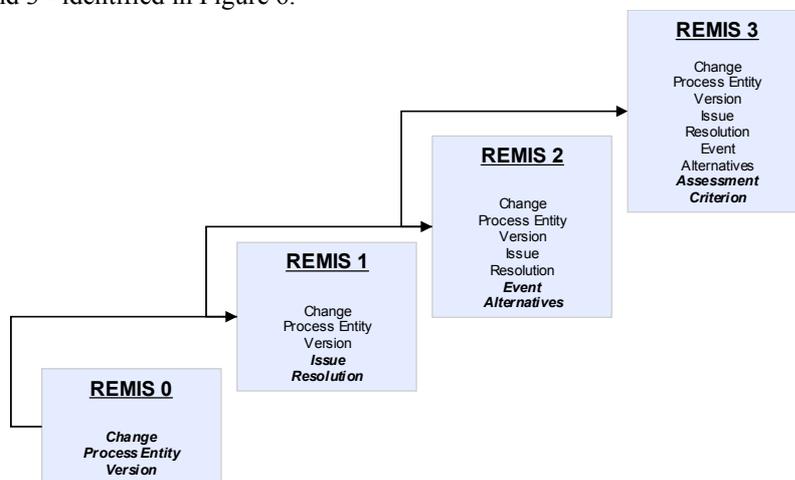

**Figure 6. Incremental Strategy**

The purpose of the REMIS 0 level is to capture the basic justification for changes to process model entities that belong to a given model version. At this level, rationale information only consists of the justification for a change. This can be a short description of the reason for performing a change.

The purpose of the REMIS 1 level is to capture the basic structuring of the reasoning behind a decision. At this level, rationale information consists of the issues and the respective resolutions that generate changes. This information can be found in organizations that use any sort of problem/resolution management process. Usually such processes are supported by a bug tracking systems where this information is captured [13].

The purpose of the REMIS 2 level is to capture the elaborated reasoning of a decision. At this level, rationale information consists of the events, the issues and their respective alternatives, and the resolutions that generate changes. Information about the alternatives cannot be found in organizations that use typical problem/resolution processes/tools, because they are not equipped to collect this kind of information. This is the reason why collection of this information is optional. However, the collection of alternatives is important for organizations because they reveal the style or preferences of the teams in charge of evolving the model. Alternatives that were not taken into account are especially important in those cases where knowledge about the application domain is minimal because the description of these alternatives offers the opportunity to retrospectively consider what should be or what should not be done in the future.

The purpose of the REMIS 3 level consists of understanding the influence of criteria on a decision. This is the highest level. In it, the most comprehensive rationale information is collected. Rationale information consists of the events, the issues, the respective alternatives, the criteria taken into account for assessing alternatives, and, finally, the resolutions that generate changes. Eliciting criteria and assessing them are optional activities.

Definition of the criteria varies from project to project. External definitions of criteria can also be incorporated into the project definition.

## 6. Fulfillment of the Requirements

The focus – its underlying rationale – for the REMIS approach is upon satisfying requirement R4 (Allow incremental adoption of the guidelines, tools and techniques). Four levels of change information capture and application are described in Section 5. These allow organizations to initially make a minimal investment in, and incur a minimal impact for, tracking changes to a system so that the purpose of individual changes may be explained and argued, and previously considered, but rejected, changes may be effectively and efficiently re-considered. As an organization's needs and capabilities to track changes increase, and its willingness to incur the impacts increases, the organization may move to more expansive 'levels' of the REMIS approach. The levels are defined to support the gradual and smooth introduction of capability as it has been observed in practice.

Unlike previous rationale conceptual models [5], the REMIS conceptual model is defined to allow incremental expansion of attention to information from, first, basic information regarding the changes made at level REMIS 0 to, ultimately at level REMIS 3, information regarding not only the changes but also the events precipitating them, the alternative changes that were considered, and the rationale underlying the choice of the change that was made. This depiction emphasizes the fact that the underlying conceptual model allows 'expansion upon demand', in other words:

expansion of the conceptual model as needed to meet an organization's needs for the capture and application of change rationale information and its tolerance for the impact upon its system development efforts. The underlying conceptual model therefore not only satisfies requirement R1 but also satisfies this requirement with a conceptual model which is considerably better – more flexible and incremental – than previously-developed models.

The REMIS approach also satisfies requirements R2 and R3. It provides techniques and supporting tools that support an organization's capture and application of process model rationale information. These techniques and tools have been defined as a result of several exercises in a variety of industrial projects. They are available to organizations which have an interest in applying the REMIS approach to rationale capture and application. And they will evolve through their future application to various situations.

## 7. Summary and Outlook

This article presents an approach - based on requirements that were derived from observing development and maintenance practices in industry - to incrementally deploying process model rationale support. In addition, the underlying REMIS approach is described that consists of a flexible conceptual model and an associated method, both supporting the effective and efficient collection and application of information about a process model's design alternatives and their selection rationale. REMIS is based on several extensive software process change exercises in industry.

Summarizing our experience with deploying rationale support we have observed that organizations should deploy rationale concepts incrementally and that this deployment process might take quite long (up to several years). The approach described in this article can be seen as a good basis and applications of the approach indicate that rationale support provides significant contributions to the expected higher-level benefits (such as reduction of evolution cost).

Based on experience with developing the method and introducing it in industry. several open questions and research directions have been identified. A selection of these topics that might be subject to future work is the following:
− What are suitable techniques for integrating and aggregating rationales to provide support for higher-level understanding and decision making?
− How to visualize the history of process models in a way that the history can be easily explained with the help of the rationale?
− How to demonstrate the value of rationale support to the higher-level goals of an organization?

**Acknowledgments**. We would to thank Sonnhild Namingha from Fraunhofer IESE for proofreading this paper.